\documentclass[preprint]{aastex}
\usepackage{graphicx,amsmath,color,natbib}

\newcommand{\metal}{\mbox{[Fe/H]}}

\newcommand{\vlos}{v_{\rm{los}}}
\newcommand{\kms}{\,\rm{km\,s^{-1}}}
\newcommand{\kpc}{\,\rm{kpc}}

\newcommand{\vhalo}{v_{\rm{halo}}}
\newcommand{\vsun}{\mathbf{v}_{\odot}}
\newcommand{\sigmaloshat}{\hat{\sigma}_{\rm{los}}}
\newcommand{\loglike}{\mathcal{L}}
\newcommand{\vlsr}{\mathbf{v}_{\rm{LSR}}}
\newcommand{\scriptr}{r}

\newcommand{\sigmalos}{\sigma_{\rm{los}}}

\newcommand{\nbhbbr}{733}               
\newcommand{\nbhbfaint}{437}            
\newcommand{\nbhb}{1170}                
\newcommand{\nbhbverybright}{227}       
\newcommand{\vysolar}{245.9}            
\newcommand{\vhaloval}{23.8}            
\newcommand{\vysolarminushalo}{222.1}   
\newcommand{\vysolarminushaloerr}{7.7}  
\newcommand{\sigmar}{101.4}             
\newcommand{\sigmatheta}{97.7}          
\newcommand{\sigmaphi}{107.4}           
\newcommand{\sigmalosval}{101.6}        
\newcommand{\vysolarerror}{13.5}        
\newcommand{\vhaloerror}{20.1}          
\newcommand{\sigmarerror}{2.8}          
\newcommand{\sigmathetaerror}{16.4}     
\newcommand{\sigmaphierror}{16.6}       
\newcommand{\sigmaloserror}{3.0}        

\begin{document}

\title{Blue horizontal branch stars in the Sloan Digital Sky Survey:\\
	II. Kinematics of the Galactic halo}
\author{
Edwin~Sirko\altaffilmark{\ref{princeton}},
Jeremy~Goodman\altaffilmark{\ref{princeton}},
Gillian~R.~Knapp\altaffilmark{\ref{princeton}},
Jon~Brinkmann\altaffilmark{\ref{apo}},
{\v Z}eljko~Ivezi\'c\altaffilmark{\ref{princeton}},
Edwin~J.~Knerr\altaffilmark{\ref{princeton},\ref{minnesota}}
David~Schlegel\altaffilmark{\ref{princeton}},
Donald~P.~Schneider\altaffilmark{\ref{pennstate}},
Donald~G.~York\altaffilmark{\ref{chicago}}
}
\altaffiltext{1}{Princeton University Observatory, Princeton,
	NJ 08544\label{princeton}}
\altaffiltext{2}{Apache Point Observatory, P.O. Box 59, Sunspot, 
	NM 88349\label{apo}}
\altaffiltext{3}{Law School, University of Minnesota, 
	Minneapolis, MN 55455\label{minnesota}}
\altaffiltext{4}{Department of Astronomy and Astrophysics, 
	the Pennsylvania State University, University Park, 
	PA 16802\label{pennstate}}
\altaffiltext{5}{Department of Astronomy and Astrophysics, 
	University of Chicago, 5640 South Ellis Avenue, 
	Chicago, IL 60637\label{chicago}}

\begin{abstract}

We carry out a maximum-likelihood kinematic analysis of
a sample of $\nbhb$ blue
horizontal branch (BHB) stars from the Sloan Digital Sky Survey
presented in \citet{halo1} (Paper~I).  Monte Carlo simulations and
resampling show that the results are robust to
distance and velocity errors at least as large as the estimated
errors from Paper~I.  The best-fit
velocities of the Sun (circular) and halo (rotational)
are $\vysolar \pm
\vysolarerror \kms$ and $\vhaloval \pm \vhaloerror \kms$ but are
strongly covariant, so that $v_{\odot} - \vhalo = \vysolarminushalo
\pm \vysolarminushaloerr \kms$.  
If one adopts standard values for the local standard of
rest and solar motion, then the halo scarcely rotates.  The velocity
ellipsoid inferred for our sample is much more isotropic
[$(\sigma_r,\sigma_\theta,\sigma_\phi) =(\sigmar \pm \sigmarerror,
\sigmatheta \pm \sigmathetaerror, \sigmaphi \pm \sigmaphierror) \kms$]
than that of halo stars in the solar neighborhood, in agreement with a
recent study of the distant halo by \citet{Sommer-Larsen_etal97}.
The line-of-sight velocity distribution of the entire sample, corrected
for the Sun's motion, is
accurately gaussian with a dispersion of $\sigmalosval \pm
\sigmaloserror \kms$.

\end{abstract}



\section{Introduction}

Among all visible components of the Galaxy, the stellar halo is
the oldest and occupies the largest volume.
It is therefore of interest both as a record of the early history
of the Galaxy and as a tracer of the Galactic potential.

The kinematics of halo stars in the solar neighborhood are
well established, as shown in Table~\ref{t:localhalo}.
For present purposes, the solar neighborhood extends as far
as a few kpc from the Sun, but to a distance less than
the Sun's own Galactocentric radius.
Evidently, the systemic rotation $\vhalo$ of the local
halo is small or zero, and the velocity ellipsoid
is distinctly elongated in the radial direction.

The more distant halo has been
less well studied.  Proper motions are generally not
measurable for distant stars, so that line-of-sight velocities
must be analyzed under \emph{a priori}
assumptions about the symmetries of the velocity ellipsoid.
The stars are also fainter, reducing the accuracy of the measured quantities.

Much work on the distant halo has been done using globular clusters,
which have the advantage of being
distinctive and bright, but have important drawbacks: there are
only $\sim 10^2$ of them, so that statistical uncertainties are large,
and special conditions of formation or selective destruction by
Galactic tides may have biased the properties of the cluster system
compared to those of the field.  \citet{Frenk_White80} analyzed the
kinematics of a sample of 66 globular clusters under the assumption
that the systemic rotation $\vhalo$ is constant; they
concluded that the residual velocities are consistent with being
isotropic within broad uncertainties.  For a subsample of 44
metal-poor (``F'') clusters, they found $\vhalo=42\pm33\kms$ 
and a root-mean-square (rms) residual
line-of-sight velocity $\sigmalos=124\kms$.  In a similar
analysis of clusters with $\mbox{[Fe/H]}<-0.8$, \citet{Armandroff89}
found $\vhalo=46\pm29\kms$ and $\sigmalos
=116\pm10\kms$.  The restriction to low metallicities follows
the discovery by \citet{Zinn85} that the more metal-rich clusters have
a distribution and kinematics intermediate between those of the
stellar halo and thin disk.

\citet{Ratnatunga_Freeman89} sampled the halo field using relatively
faint red giants at high Galactic latitude.  Their summary for the
more metal-poor stars near the south Galactic pole concludes that
$\sigma_z$ is approximately constant at $\sim 75\kms$ out to $25\kpc$,
which would be consistent with the solar neighborhood
(Table~\ref{t:localhalo}) if the velocity ellipsoid is constant in
cylindrical coordinates; however, for their most distant subsample
near the SGP ($\langle D\rangle=17.1\kpc$), they measured $\sigmalos
=42\pm10$ (10 stars).

In a study similar to that reported here, 
\citet{Sommer-Larsen_etal97} analyzed a
sample of 679 blue horizontal branch stars drawn from multiple sources
and spanning Galactocentric radii $\sim 7-65\kpc$.  With line-of-sight
velocities and photometric distances, they found that the radial
velocity dispersion (in spherical coordinates) falls rather sharply
from its value in the solar neighborhood, $\sigma_r=140\pm10\kms$, to
an asymptotic value $89\pm 19\kms$ at $r\gtrsim 20\kpc$.  By requiring
dynamical equilibrium in an assumed logarithmic Galactic potential,
they concluded that the tangential velocity dispersion should rise
from its local value to $\sigma_\theta\approx\sigma_\phi\approx
137\kms$ at large $r$.  They did not claim, however, to have detected
this rise directly in their data.  They further suggested that the
shift from radial to tangential anisotropy with increasing 
Galactocentric radius
might support theories in which the halo formed by heirarchical
merging of smaller systems.

Because of the differences in kinematics cited above,
it appears that local high-velocity, low-metallicity stars may
not be entirely representative of the more distant halo.
To study the more distant halo, 
we have isolated a sample of $\nbhb$ blue horizontal branch (BHB) 
stars from the Sloan 
Digital Sky Survey \citep[][hereafter Paper I]{halo1}.  In that paper,
we estimated contamination from foreground stars and blue stragglers to
be less than $10\%$ for bright stars ($g<18$) and about $25\%$ for faint
stars ($g>18$).  An improved color-absolute magnitude relation was
also derived, and these absolute magnitudes are used throughout
this paper to derive distance moduli.
This sample of BHB stars provides effective tracers of the Galactic 
halo, with no kinematic bias.  The sample may have selection bias, and
is probably far from complete, but these biases do not affect the
kinematic analyses presented in this paper other than
degrading the statistics and increasing the errors.
Using maximum-likelihood methods, we fit for the mean
rotation and principle axes of the halo velocity ellipsoid on the
assumption that these are constant in spherical or cylindrical
coordinates.  We also fit for the orbital velocity of the Sun.

The present paper is organized as follows.
We first discuss an overview of the kinematic analysis in
\S\ref{s:ve_overview} and appendix~\ref{s:maxlike}.
The results for the kinematic parameters of the halo are presented
in \S\ref{s:ve_results} and summarized in Table~\ref{t:results}.
Monte Carlo simulations which validate the analysis are presented
in \S\ref{s:montecarlo}.
\S\ref{s:discussion} summarizes our main conclusions and discusses
the agenda for future work.

\subsection{Terminology used in this paper}\label{s:terminology}
The following terminology and conventions are used throughout this work.
A heliocentric right-handed Cartesian coordinate system 
$(x,y,z)$ is oriented so that $+x$ points towards the 
Galactic center (GC) and $+z$ 
points towards the North Galactic Pole, such that $+y$ is the direction
of disk rotation.  Thus, $\vsun = (v_{\odot,x},v_{\odot,y},v_{\odot,z}) = 
\vlsr + (U,V,W)$ where $(U,V,W)$ is the solar motion with respect
to the local standard of rest (LSR).  
The velocity of the LSR is taken to be $\vlsr = (0,220,0) \kms$ 
\citep{kerr_lyndenbell_1986}, and the canonical solar motion
vector is $(U,V,W) = (10,5,7) \kms$ 
\citep[rounded to integer values]{dehnen_binney_1998}.
Galactocentric spherical $(r,\theta,\phi)$
and cylindrical $(R,\phi,z)$ coordinate systems will also be used.
We make a rudimentary model of halo bulk rotation parametrized by
a single velocity $\vhalo$.
A positive $\vhalo$ indicates a flat halo rotation curve in the same
sense as the disk rotation.
Radial velocities to stars are measured as redshifts which
are subsequently and immediately corrected for 
the Earth's motion around the Sun and 
the Earth's rotation about its own axis.  This heliocentric radial velocity
$v_r$ contains a component due to the solar velocity 
$\vsun$.\footnote{Here the subscript ``$r$'' is used for components
along the line of sight, but elsewhere it will be used for the radial
component in spherical coordinates based on the Galactic Center.  The
context should make our meaning clear.}  
The ``line-of-sight'' velocity $\vlos$
will be defined as $v_r$ plus the projection of $\vsun$ onto the
line-of-sight to the star, minus the projection of its expected
rotation component $-\vhalo \boldsymbol{\hat{\phi}}$ (if any).  That is, 
$\vlos$ is the component of the purely random component of velocity
in a model of the halo with a flat rotation curve, projected onto
the line of sight.  We find $v_{\rm los}$ conceptually useful, but
one should not forget that it is contingent upon our assumptions
concerning the rotation of the Sun and the halo, whereas $v_r$
is directly observed (modulo errors).
Throughout, the distance from the Sun to the GC is assumed to be
$8 \kpc$ \citep{reid_1993}.

\section{The velocity ellipsoid of the Galactic halo: overview}
	\label{s:ve_overview}

The velocity ellipsoid is defined such that its axes align with
the principal axes of the velocity dispersion tensor, and the
lengths of its axes $\boldsymbol{\sigma}$ 
are equal to the components of the velocity
dispersion along those directions.  It is difficult to know the
orientation of the velocity ellipsoid so it is reasonable to
assume it is oriented along galactocentric spherical or cylindrical
coordinates at every point in space \citep{binney_merrifield}. 
Even with $\sim 10^3$
stars, it is difficult to assess whether and how the
components of the velocity ellipsoid vary in space, so
they are assumed constant, although we test this below by
dividing our sample into a few magnitude bins.

A maximum likelihood technique is used, as outlined in 
Appendix~\ref{s:maxlike}. 
A uniform prior is adopted \citep{lupton_1993}:
when the probability of observing the set of stars at 
the observed positions and heliocentric radial velocities 
given an assumed $\boldsymbol{\sigma}$ is maximized as a function of 
$\boldsymbol{\sigma}$, 
the probability of $\boldsymbol{\sigma}$ being the true
velocity ellipsoid is maximized.  In symbols,
$\phi(\vlos \, | \, \boldsymbol{\sigma}) \propto
	\phi(\boldsymbol{\sigma} \, | \vlos)$
where $\phi$ represents probability density.
Naturally, $\phi(\vlos \, | \, \boldsymbol{\sigma})$ depends implicitly
upon the orientation of the velocity ellipsoid
with respect to the line of sight, and hence
upon the angular position $(\alpha,\delta)$ and distance
of each star.
It is straightfoward to expand the maximum likelihood procedure to
include four additional parameters: the three components of the
velocity of the Sun with
respect to the galactic center, $\vsun$, and the bulk rotation 
velocity of the halo, $\vhalo$.

We assume that the bulk rotation 
of the halo corresponds to a uniform velocity in the 
direction of galactic rotation.
It is difficult to apply this analysis on this sample
to more complex halo rotation laws because of the moderate degeneracy
between the halo rotation and $v_{\odot,y}$ (see~figure~\ref{f:ve_conf},
discussed below); this
degeneracy would increase if one attempted to fit for more parameters
of the bulk halo rotation.

\begin{figure}
\includegraphics[width=\textwidth]{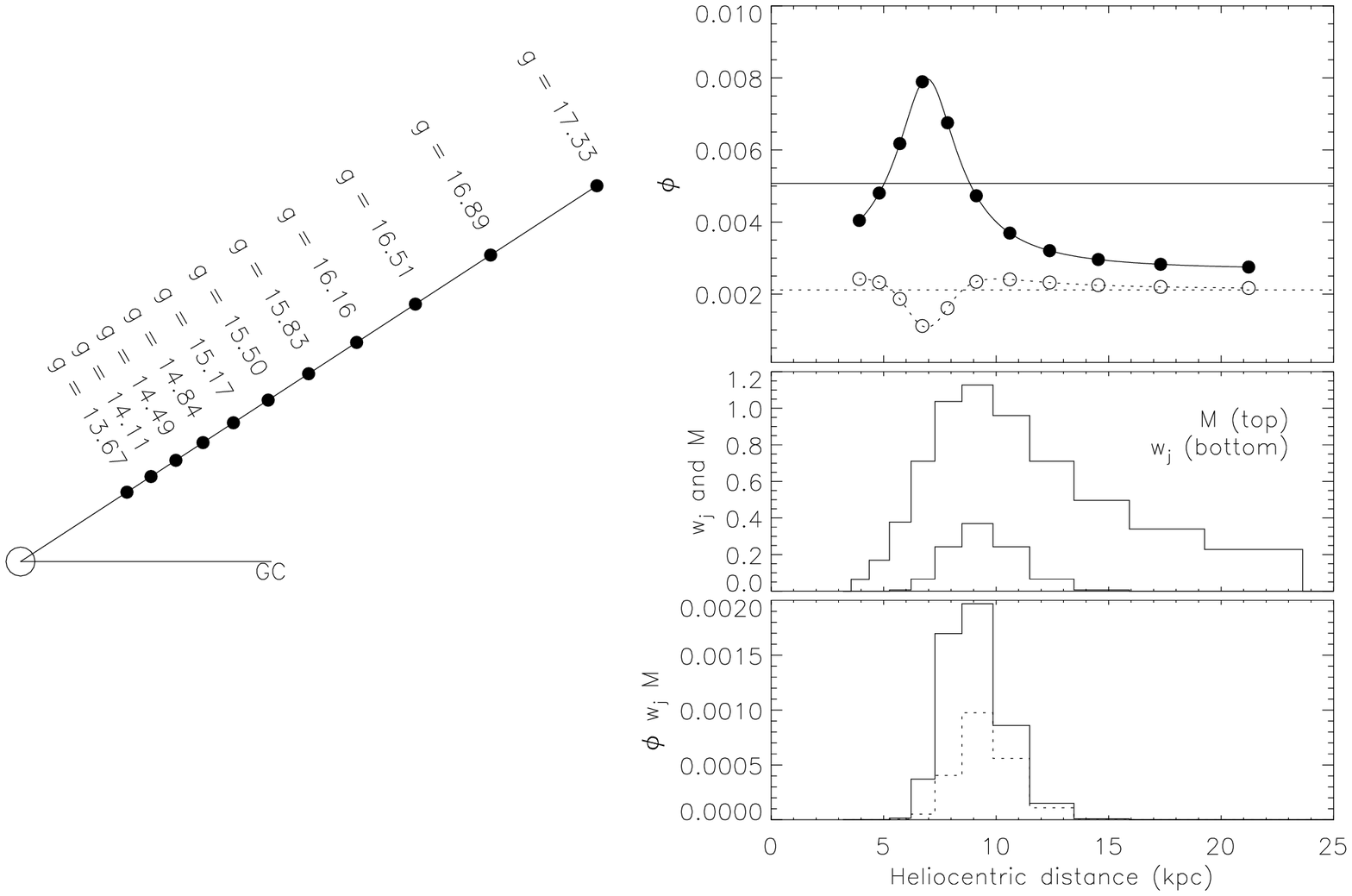}
\caption{An illustration of the technique used to incorporate
magnitude errors into the maximum likelihood analysis, for two
artificial stars.  The two stars are at the same measured position: 
$(l,b) = (0,30)$, $g = 15.5$, and absolute magnitude 0.7.  
Magnitude errors mean that
the stars could be located anywhere along the line of sight indicated
in the left panel.  One star has $\vlos = 0 \kms$
and the other has $\vlos = 100 \kms$.  Here we assign a model with
$\boldsymbol{\sigma} = (150, 50, 100) \kms $ in spherical coordinates, 
and neglect any velocity of the Sun with respect to the galactic center.
The upper right panel plots the probability density of the existence of 
each star with given $\vlos$ at the position given by the abscissa
(see equation~\protect{\ref{e:probability}}).
The solid line represents the stationary star, and the dotted line 
represents the receding star.  The middle panel shows the weights 
of Gauss-Hermite integration $w_j$ (bottom line)
and the Malmquist factor $M(r)$ (top line), as described in the text, 
which are then multiplied to $\phi$ to arrive at the solid (stationary star)
and dotted (receding star)
lines in the bottom panel.  The values of $\phi w_j M$ are summed over
the eleven abscissae, and plotted in the upper right panel as the
horizontal solid and dotted lines, for the stationary and receding stars
respectively.  }
\label{f:gh_demo}
\end{figure}

Distance moduli are derived from the color-absolute magnitude relation
of Paper~I.
The measured angular coordinates of the stars $(\alpha, \delta)$
are virtually error-free.  
However, the magnitudes, and thus distances, are susceptible to several
sources of error: extinction, photon statistics, and intrinsic 
brightness variations from star to star which create deviations
in the stars' distance moduli.
Therefore, magnitude errors should be included in the likelihood analysis.
As shown in Paper~I, typical magnitude errors are about
$\Delta m \sim 0.2$.  In anticipation of our finding,
discussed below in \S\ref{s:ve_results}, 
that the kinematic analysis is very robust with
respect to magnitude errors, we adopt an even more conservative
magnitude error: $\Delta m = 0.5$, and assume the errors follow a
gaussian distribution.
Equation~\ref{e:probability} shows the probability density $\phi$ of
observing a star at a given position in space,
line of sight velocity $\vlos$, and $\boldsymbol{\sigma}$.
Having assumed that magnitude errors have a gaussian distribution with some
width $\Delta m$, the likelihood of observing a star considering
magnitude errors can be found
by integrating the probability density of equation~\ref{e:probability}, 
multiplied by a gaussian weight function for the magnitude error, 
along the line of sight.  
Figure~\ref{f:gh_demo} shows an illustration of this procedure
for two stars with artificial data.  These two stars are constructed
to have $(l,b) = (0,30)$, $g = 15.5$, and absolute magnitude
$0.7$, which places them approximately
9 kpc from the Sun above the galactic center.  The evaluation of
the probability density $\phi$ for each star alone 
along the line of sight is shown in
the upper right panel.  One star's line-of-sight velocity is
$\vlos = 0 \kms$ (solid line and filled dots) 
and the other's is $\vlos = 100 \kms$ (dotted line and open circles). 
We apply $\boldsymbol{\sigma} = (150,50,100) \kms$ (spherical coordinates)
to this model (note that the $\phi$ component of the velocity ellipsoid
is irrelevant for this particular case).
It is easy to see several trends in the upper right plot of 
figure~\ref{f:gh_demo}.  If there were no distance measurement, for
example,
the stationary star would be most likely to be found at 
$\approx 7 \kpc$ because the line of sight coincides there with
the smallest ($\theta$) axis of the velocity ellipsoid.
In fact, the probability density of
observing a stationary star at that point is just 
$(2\pi\sigma_{\theta}^2)^{-1/2} \approx .008\,\rm{s\,km^{-1}}$, as shown.  
Also, the receding star
is least likely to be observed there because its $\vlos$ can most
easily be ``explained'' by most of its contribution coming from the
radial component of the velocity ellipsoid.  Also note that the
stationary star always has a higher probability than the 
receding star; this follows from the gaussian nature of the
probability of observing $\vlos$ in equation~\ref{e:probability}.  

To incorporate the magnitude error $\Delta m$ into the analysis,
the probability density $\phi$ as a function of distance must be multiplied
by a gaussian centered on the measured magnitude with width $\Delta m$,
and then integrated over distance.
This integration is most easily accomplished with Gauss-Hermite integration,
which samples $\phi$ at $N_{\rm p} = 11$ points, shown in the upper right
panel of figure~\ref{f:gh_demo} as dots and circles.  These 11 values
of $\phi$ at the specified abscissae are multiplied by the coresponding
weight factors $w_j$ \citep{abramowitz_stegun_1972}, 
which are plotted
in a discrete fashion as the lower line in the middle panel of 
figure~\ref{f:gh_demo}.  The consideration of magnitude errors 
also introduces Malmquist bias.  If
the number density of observable BHB stars in the Galactic halo
follows a spherical power law
$\rho \propto r^{-\alpha}$ in Galactocentric radius, then
Malmquist factor $M(r') \propto r^{\prime\,3} r^{-\alpha}$ must 
weight the probability density $\phi$, in addition to the gaussian
weight $w_j$ due to the magnitude error, where $r'$ is heliocentric distance.  
These weighting factors are normalized so that 
$\sum_j^{N_{\rm p}} w_j M(r) = 1$.
The function $M(r')$ is shown
in the middle panel of figure~\ref{f:gh_demo} as the top line,
assuming $\alpha = 3.5$. Note that for $\alpha = 3$,
$M(r \gg 8 \kpc) \to \rm{constant}$.

Finally, the probability density 
of observing a star at a given angular position,
distance modulus, and line-of-sight
velocity, with assumptions on $\boldsymbol{\sigma}$, 
$\Delta m$ and $\alpha$ for the Malmquist
term, is just $\sum_j \phi(r) w_j M(r)$.  The term 
$\phi(r) w_j M(r)$ is plotted in the lower panel of figure~\ref{f:gh_demo}
for our two fictitious stars; again the stationary star is the solid line
and the receding star is the dotted line.  When these terms are 
summed over $j$, we arrive at new determinations for the probability density
of observing the star, indicated in the upper right panel of 
figure~\ref{f:gh_demo} by the horizontal lines.  The original
estimate of $\phi$ was given by the middle dot and circle (at 
$\approx 9 \kpc$), so it is evident that the ``hill'' and the ``dip''
in $\phi(r)$ for the two stars are responsible for modifying
the original estimate of $\phi$ up or down
in the correct direction.

The quantity $\alpha$ for the halo is 
usually taken to be between about 3 and 3.5 
\citep[and references therein]{yanny_etal_2000}.
In the kinematic analysis below we assume $\alpha = 3.5$, following
the study of horizontal branch stars of \citet{kinman_suntzeff_kraft_1994}.
However, we find our results are fairly insensitive
to $\alpha$; even $\alpha = 10$ results only in
small changes to our final results.  

Another possible source of measurement error is in the radial velocities.
As determined in Paper~I, typical velocity errors may
be as high as $\sim 30 \kms$.  According to equation~\ref{e:probability},
the probability density of observing a star at a certain line-of-sight velocity
$\vlos$ (with respect to the GC) follows a gaussian distribution,
centered on $\vlos = 0 \kms$,
with a dispersion $\sigmaloshat$ 
that depends on geometry and $\boldsymbol{\sigma}$.
Thus the probability density of observing the star at $\vlos$
considering a possible velocity error that follows a gaussian 
distribution with dispersion $\Delta v$ is just
\begin{eqnarray}\label{e:verror}
\phi_{\rm{e}}(\vlos) &=& \frac{1}{\mathit{\Gamma} \sqrt{2\pi \Delta v^2}}
	\int_{-\infty}^{\infty} \exp{ \left(\frac{-v^2}{2 \sigmaloshat^2} - 
		\frac{(v - \vlos)^2}{2 \Delta v^2} \right) } dv \nonumber \\
&=& \frac{1}{\mathit{\Gamma} \sqrt{1 + (\Delta v/\sigmaloshat)^2}} 
	\exp{ \left( \frac{-\vlos^2}{2(\sigmaloshat^2 + \Delta v^2)} \right) };
\end{eqnarray}
note this reduces to the original equation~\ref{e:probability} as
$\Delta v \to 0$.

We have discussed the various modifications available to the maximum
likelihood technique.  In summary, they are the following:
\begin{itemize}
\item Subsample. Although we first determine the velocity ellipsoid
using all stars, we can extract subsamples, specifically in certain
magnitude bins, to investigate whether the velocity ellipsoid is a function
of position in space.
\item Spherical or cylindrical coordinates.  The choice affects the values
of $\mathbf{x}$ in the maximum likelihood procedure but is otherwise
straightfoward.
\item Magnitude errors.  We either ignore magnitude errors or else use
11-point Gauss-Hermite integration with $\Delta m = 0.5$ and $\alpha = 3.5$.
\item Velocity errors.  We either ignore velocity errors or else use
equation~\ref{e:verror} with $\Delta v = 30 \kms$.
\item Number of fitting parameters.  Several combinations of fitting
parameters are used.  We fit the three $\boldsymbol{\sigma}$ parameters
keeping $(\vsun, \vhalo)$ at canonical values; we fit 
$(\vhalo,\boldsymbol{\sigma})$ keeping $\vsun$ 
at its canonical values; and we also fit
all seven parameters simultaneously. 
\end{itemize}

\section{Results}\label{s:ve_results}

Table~\ref{t:results} contains the determinations of 
$(\vsun, \vhalo, \boldsymbol{\sigma})$ for several
combinations of model assumptions discussed above; the first
four columns describe these model assumptions.  
The first column states what criteria were used to select 
subsets of BHB stars from the total set of $\nbhb$ BHB stars, 
and the number of stars in each subset is listed in the second column.
The third column indicates whether a cylindrical or spherical geometry
was assumed.  The fourth column indicates how errors were handled;
``0'' indicates that errors were not considered.  As described in
\S\ref{s:ve_overview}, ``m'' indicates
that 11-point Gauss-Hermite integration was used with $\Delta m = 0.5$, 
and ``v'' indicates that
velocity errors were considered with $\Delta v = 30$.
The following columns show the parameter
determinations.  
All determinations in table~\ref{t:results} are for the three-parameter
fit with $(\vsun,\vhalo)$ fixed at canonical values 
\citep{dehnen_binney_1998} as shown, except
for the second and third lines which fit four and seven parameters, 
respectively.
The quoted errors consider 
correlations among the parameters, and the errors in parentheses 
do not consider the correlations.  The former error is defined as 
$\sqrt{-W_i/2}$ and the latter as $1/\sqrt{-2M_i}$ where $M_i$ represents
the diagonal elements of the second-derivative matrix of the log-likelihood
function, and $W_i$ represents the diagonal elements of the inverse of
that matrix.

The first row in Table~\ref{t:results}
 shows the results for the velocity ellipsoid
for the complete sample ($N_{\rm{stars}} = \nbhb$) in spherical 
coordinates, with no magnitude or velocity errors considered.
The Sun's velocity was fixed at the canonical value of
$\vsun = \vlsr + (U,V,W) = (0,220,0) + (10,5,7) \kms$ 
\citep{dehnen_binney_1998}.
Under these assumptions all three components of the velocity
ellipsoid are about $100 \kms$.
The second row in Table~\ref{t:results} shows the results of 
simultaneously fitting $\vhalo$ and $\boldsymbol{\sigma}$
to the 
same sample and otherwise same set of assumptions discussed above.
The velocity ellipsoid has essentially the same determined values,
and the rotation velocity of the halo is consistent with zero.
The third row in Table~\ref{t:results}
is the seven parameter fit to the sample.
Again, the velocity ellipsoid is essentially the same,
and the rotation velocity of the halo is seen to be marginally consistent
with zero.
Reassuringly, the velocity of the Sun with respect to the galactic
halo in the $y$ direction 
(or equivalently, the Sun's velocity in an inertial frame
assuming the halo is actually non-rotating)
is close to the expected value:
$v_{\odot,y}-\vhalo = \vysolarminushalo \pm \vysolarminushaloerr \kms$.  
The error follows from rotating the eigenvectors by $45\deg$ and is smaller
than either error for $v_{\odot,y}$ or $\vhalo$ because the two
quantities are strongly covariant.
The $x$ and $z$ directions of
$\vsun$ are new measurements of the solar motion $(U,W)$.  Our value
of $U$ is consistent with that of \citet{dehnen_binney_1998}; however
our value of $W$ differs by several $\sigma$.

\begin{figure}
\includegraphics[width=\textwidth]{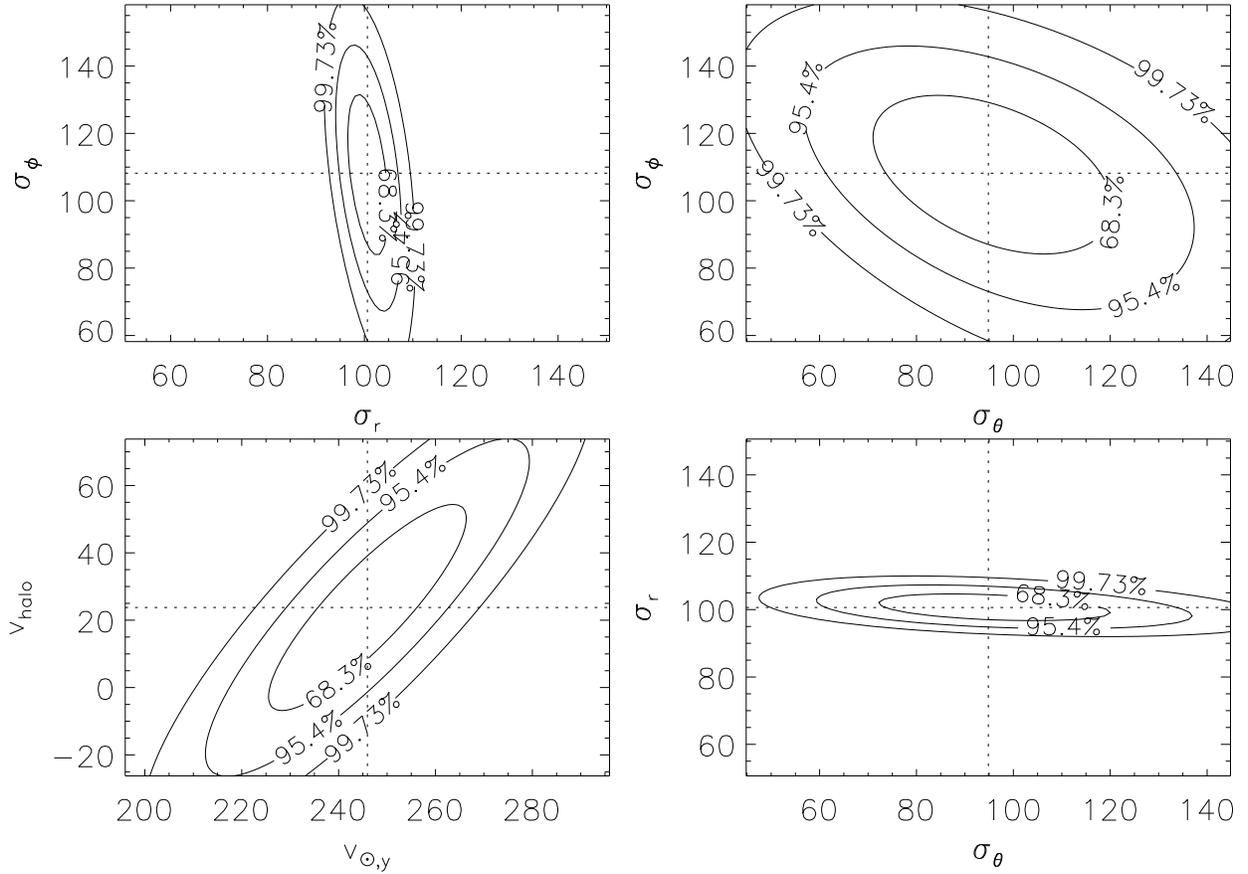}
\caption{Approximate confidence regions for the components of the
velocity ellipsoid determined from the complete BHB sample ($\nbhb$ stars)
in spherical coordinates, demonstrating $\sigma_r$ is very well
constrained and that each component is anticorrelated with the other two.
In the lower left panel, the confidence regions for the halo rotation
and the solar velocity in the $+y$ direction are shown.}
\label{f:ve_conf}
\end{figure}

Figure~\ref{f:ve_conf} shows confidence regions for 
the components of the velocity ellipsoid for this case.
Each panel in figure~\ref{f:ve_conf} is a two-dimensional
parameter space where the other five parameters are held at their
determined values and assumed to have symmetric errors, so the
confidence regions are only approximatations to those one would obtain
by integrating over the other five dimensions.
Note that $\sigma_r$ is very well constrained. This is because the
typical distance to the halo stars in our sample is
large compared to the distance
from the Sun to the galactic center, so measurements of $\vlos$
from Earth very nearly align with direct measurements of 
the galactocentric radial component of velocity.
Also, each component of the velocity ellipsoid is anticorrelated
with the other two; this demonstrates an intuitive ``conservation law'' 
of velocities.  In the lower left panel, it is apparent that $\vhalo$
and $v_{\odot,y}$ are correlated, so that the quantity $v_{\odot,y} - \vhalo$
is more tightly constrained than either quantity alone.  This of course
follows from the reasonable assumption that the halo rotation 
at the location of the Sun is in the $y$ direction.

The result that $\sigma_r \approx 100 \kms$ contrasts with
most estimates of the velocity ellipsoid based on stellar samples
in the solar neighborhood \citep{woolley_1978,pier_1984,
carney_latham_1986,norris_1986,layden_etal_1996}.  We now investigate
some other sets of assumptions that may influence the results.

The fourth line in Table~\ref{t:results}
introduces magnitude errors into the 
set of assumptions.  These errors, at least in
our implementation of a gaussian with width $\Delta m = 0.5$, apparently
have little effect on the final results.  
Note that for a given kinematic model,
distance affects the distribution of $\vlos$ only
through the orientation of the velocity ellipsoid with respect to
the line of sight. If the dispersions were isotropic and the
halo nonrotating, which seem to be nearly the case,
then $\vlos$ would be independent of distance.  This may explain
the insensitivity to magnitude errors.
The fifth line in Table~\ref{t:results}
incorporates both magnitude and velocity errors.
Again, the final results are seen to remain relatively constant;
however the values of the velocity ellipsoid components decrease
modestly.  In light of equation~\ref{e:probability}, the
probability of observing a star at a certain velocity $\vlos$ is
always higher for smaller values of $|\vlos|$; thus velocity errors
tend to ``favor'' smaller velocities, which propagate to the smaller
values of the velocity ellipsoid.  However, the effect is small, 
of order $\sqrt{\sigma^2 + \Delta v^2} - \sigma \approx 5 \kms$.

The next three lines in Table~\ref{t:results}
show three-parameter
determinations for the complete sample of BHB stars assuming
the velocity ellipsoid is oriented along cylindrical coordinates.
As for the spherical coordinates case, all three components of the
velocity ellipsoid are still about $100 \kms$.  

Determinations for cylindrical scenarios incorporating magnitude and velocity
errors are shown in the tables, and as before, 
do not appear to affect the final results appreciably.  
Since velocity errors
behave in a predictable way and have little effect on our final results,
and magnitude errors have no noticable effect, we do not consider 
these errors further.
It should be remembered that our error estimate
was based on a set of highly simplified, and perhaps dubious,
assumptions that do not take into account systematic errors, for instance.

In the following rows in Table~\ref{t:results}
 the sample of stars is modified.
There are $\nbhbbr$ stars with $g < 18$, so these stars
extend from $\sim 7 \kpc$ to $\sim 30 \kpc$.  The sample
with $g < 16$
contains the $\nbhbverybright$ brightest stars (distances out to
$11.5 \kpc$).  The velocity ellipsoid is seen to remain remarkably
constant down to distances from the Sun only marginally outside
the coverage of previous surveys 
(e.g., $V \lesssim 13$ for the sample of \citet{layden_etal_1996}).  
The most intriguing
result is that the radial component of the velocity ellipsoid 
$\sigma_r$ (or $\sigma_R$ in cylindrical coordinates) just
outside the solar neighborhood is significantly
smaller than the canonical value of $\sim 150 \kms$ in the solar
neighborhood (Table~\ref{t:localhalo}).  
This incompatibility will be further discussed in
\S\ref{s:discussion}. 

The sample with $g > 18$ contains the 
$\nbhbfaint$ faintest stars. 
Since the stars are all very far away,
the non-radial components of 
the velocity ellipsoid are very poorly constrained.  Note that
the quoted errors may also be ill-determined because of departures
from gaussianity for small eigenvalues. 
However, the essential result remains the same: the radial component of the
velocity ellipsoid is $\sim 100 \kms$.

Finally, to address concerns that the determination of the velocity
ellipsoid may be contaminated by halo substructure, the last two lines
of Table~\ref{t:results} show the results for all stars excluding the
Sagittarius stream.  This exclusion is conservatively accomplished by
simply rejecting all BHB stars within 10 degrees of the great circle with
pole $(\ell, b) = (94,11)$, which approximates the Sagittarius stream
\citep[][Paper~I]{ivezic_etal_2003}.
Since the
Sagittarius stream is the most conspicuous substructure in our sample
of BHB stars (Paper~I), the stability of the velocity ellipsoid solution
indicates that halo substructure does not appreciably influence the
results.

\section{Monte Carlo simulations}\label{s:montecarlo}

An independent way of verifying the accuracy of the maximum likelihood
technique is through Monte Carlo simulations.  In case the geometry
of the SDSS footprint (see Paper~I) affects the analysis,
we assemble $\nbhb$ artificial stars 
with the same angular coordinates and measured $g$ magnitudes
as the complete BHB sample, but with newly constructed velocities.
For simplicity absolute magnitudes are assumed constant at $0.7$.
Only velocity ellipsoids aligned with spherical coordinates will
be considered here.  These Monte Carlo simulations perform the three-parameter
fit to the velocity ellipsoid with $\vsun = (10,225,7) \kms$
and $\vhalo = 0 \kms$.  Note, however, that the values of $\vsun$ and
$\vhalo$ are irrelevant here since they are first added to the stars'
velocities in construction, and then subtracted as known
in the parameter determination.

We run 100 Monte Carlo simulations on each of four setups.  The first
setup does not consider magnitude or velocity errors; the second
setup allows for $\Delta m = 0.5$; the third assumes $\Delta v = 30 \kms$.  
The fourth setup considers
the magnitude and velocity errors as well as allowing for
contamination.

The constructed set of artificial stars has known space positions,
so it is straightforward to apply an assumed velocity ellipsoid
to obtain a simulated three-dimensional space velocity for each star.
Here we assume $\boldsymbol{\sigma} = (150,100,80) \kms$ 
to strain the maximum likelihood analysis.
The three-dimensional velocities are then projected onto the line
of sight to each star to obtain $\vlos$, and then the projections
of $\vsun$ and $\vhalo$ are subtracted to obtain $v_r$.  
This comprises the simulated data set for the first setup.  To incorporate
magnitude errors, the radial distances to the stars are shuffled,
i.e. the $g$ magnitudes are shifted by a gaussian error distribution
of width $\Delta m$, $v_r$ is calculated as before, and then the 
$g$ magnitudes are shifted back to their ``observed'' positions.
Velocity errors are incorporated by adding an error to each observed
$v_r$.  To incorporate contamination, a fraction of
stars are randomly flagged as disk stars.  These stars are then
assumed to be 2.5 times closer (2 mag) to the Sun than they would be if they
were BHB stars. 
Their velocities are taken to be $220 \kms$ in the $-\phi$ direction
plus three random components from a \emph{disk} velocity dispersion,
which is assumed isotropic with components $(20,20,20) \kms$.
\citet{dehnen_binney_1998} find a non-isotropic disk
velocity dispersion with components of the same order
(depending on color), but our intent is to 
contaminate the simulated halo star population in a simple and 
controlled way. 

\begin{figure}
\includegraphics[width=\textwidth]{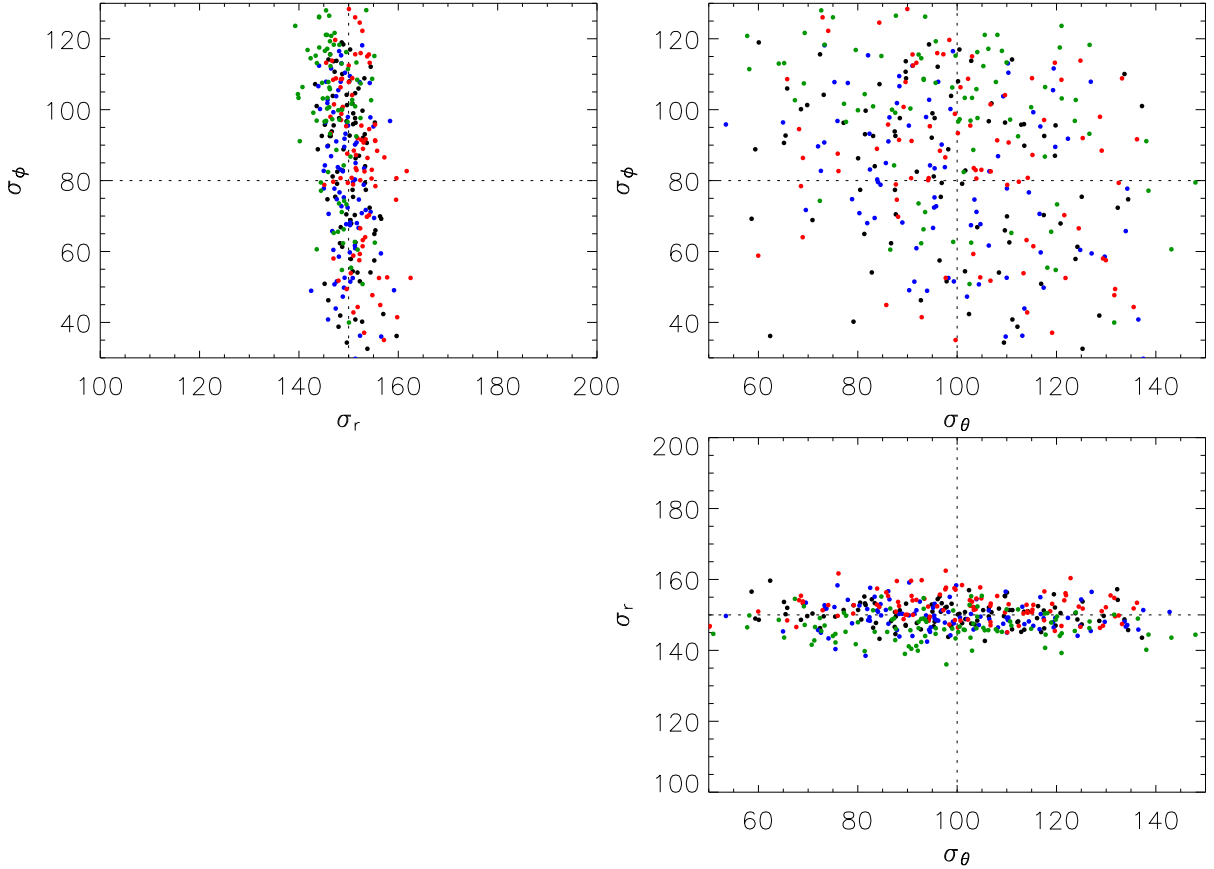}
\caption{Each dot represents the results from a Monte Carlo simulation
with input $\boldsymbol{\sigma} = (150,100,80) \kms$ in spherical
coordinates.  Black dots assume no magnitude or velocity errors.
Blue dots assume the stars are not really at the heliocentric radial
positions observed, with an error $\Delta m = 0.5$.  Red dots
assume the positions are accurate, but that the velocities are 
measured with an error $\Delta v = 30 \kms$.  Green dots assume
magnitude and velocity errors, as well as a 10\% contamination
from disk stars.  This plot provides evidence that the error
estimate as shown in figure~\protect\ref{f:ve_conf} is accurate.
Additionally, the radial component of the velocity ellipsoid is
very well constrained, as in the analysis of the real data.
One can also make out slight anticorrelations between all three 
parameters, as with the real data.  Magnitude errors (blue dots)
have no perceivable effect on the parameter determinations of the
velocity ellipsoid.  Velocity errors of $30 \kms$ (red dots)
introduce a small positive shift of order $\sim 5 \kms$
to each of the
three components of the velocity ellipsoid, which is most easily
seen in the radial component.  Contamination (green dots) 
shifts $\sigma_{\theta}$ and $\sigma_r$ to lesser values and 
$\sigma_{\phi}$ to larger values, as expected; also see 
Table~\protect{\ref{t:mc_meanrms}}.}
\label{f:ve_mc}
\end{figure}

Figure~\ref{f:ve_mc} shows the results of these Monte Carlo simulations
with the four setups colored black (no errors), blue (magnitude
errors), red (velocity errors), and green (magnitude and velocity 
errors with 10\% contamination).  As shown, the scatter in the
results of the Monte Carlo simulations is comparable to the confidence regions
shown in figure~\ref{f:ve_conf}, so our error analysis is verified
to be reasonable.  As in the analysis of the real data, the radial
component of the velocity ellipsoid $\sigma_r$ is very well constrained;
one can also see anticorrelation between any two components.
The black dots and blue dots occupy essentially the same region of 
parameter space; this verifies, independently of the method used
in \S\ref{s:ve_results}, that magnitude errors of the order
$\Delta m$ do not affect our results significantly.  

As shown in Table~\ref{t:mc_meanrms}, the mean determinations of 
$\boldsymbol{\sigma}$ over 100 Monte Carlo simulations recover the
input values for the ``no errors'' case and the ``magnitude errors only''
case.  
It is also clear that velocity errors of scale $\Delta v = 30 \kms$ push
the measured values higher by $\sim 5 \kms$ because
the error adds in quadrature.  This result is
consistent with the independent determination in \S\ref{s:ve_results}
which stated that with $\Delta v = 30 \kms$ the ``true'' values
of the velocity ellipsoid
(i.e., those measured using the analysis that allows for velocity errors)
were $\sim 5 \kms$ lower than the ``measured'' values 
(not considering velocity errors).
Finally, it is interesting that for the case which includes contamination
by disk stars, the determination of $\sigma_{\phi}$ increases while
those of $\sigma_r$ and $\sigma_{\theta}$ both decrease.
This is because the disk stars are constructed to travel primarily 
in the $\phi$ direction.  

However, the main conclusion
to be gained from these Monte Carlo simulations is that the radial
component of the velocity ellipsoid never strays too far from its
correct value of $150 \kms$, even allowing for errors such as 
contamination.  Thus the maximum likelihood analysis is verified
to be reliable and stable.

\section{Summary and discussion}\label{s:discussion}

Applying maximum-likelihood methods to the line-of-sight velocities
and photometric distances of BHB stars, we have modeled the kinematics of the
relatively distant galactic halo.  (The median distance of the sample
is about $25\kpc$.)
No significant anisotropy of the velocity ellipsoid is detected.  In a
model aligned with spherical coordinates, 
for example, $\boldsymbol{\sigma} = (\sigmar,\sigmatheta,\sigmaphi) \kms$.
However, the errors in
$\sigma_\theta$ and $\sigma_\phi$ are much larger, $\sim 15-20\kms$,
than that of $\sigma_r$,
because of the great distances of most stars: that is, the
Galactocentric radius $\scriptr$ of most stars is more than twice that of
the Sun, so that the direction of the line of sight is not far from
radial despite the broad angular range of our sample.  This result
contrasts starkly and surprisingly with the radial anisotropy of halo
velocities in the solar neighborhood 
($\boldsymbol{\sigma} \sim (150,100,100) \kms$; see Table~\ref{t:localhalo}).
Our results are consistent, however, with the only other comparable
study of distant halo field stars: \cite{Sommer-Larsen_etal97}.

We are able to fit separately for the rotation speed of the halo
($\vhalo$) and the orbital velocity of the Sun, although the two
are strongly covariant.  The halo rotation is marginally consistent
with zero at the one-sigma level, which is $\sim 20\kms$.  The
difference 
$v_{\odot,y}-\vhalo=\vysolarminushalo\pm\ \vysolarminushaloerr \kms$ 
is more tightly constrained than either quantity separately.
This is consistent with a non-rotating halo if one accepts
the IAU recommendation for the local standard of rest,
$\vlsr=220\kms$ \citep{kerr_lyndenbell_1986} and
\citet{dehnen_binney_1998}'s
solar motion, $\vsun-\vlsr=(10,5,7)\kms$, although some authors
have advocated a substantially smaller \citep{olling_merrifield98}
or larger \citep{uemura_etal00} value of the azimuthal component
of $\vlsr$.  (If the solar rotation \emph{is} substantially less
than $225\kms$, then according to our results, the halo should
counterrotate!) 
Our result is also statistically
consistent with most of the studies cited in Table~\ref{t:localhalo},
although the latter seem to prefer a slight positive halo rotation.
Note that what those studies actually measure is the mean velocity
of local halo stars relative to the Sun; the values for
$v_{halo}$ in the fourth column of that table have been adjusted to
$V_\odot=225\kms$ so that they can be compared directly to our result.
Presumably, $\vlsr$ will soon be determined with exquisite accuracy
by a variant of the classical visual-binary method applied to stars
orbiting Sgr~$\rm A^*$, but accurate values are not in the public 
domain at the time of writing \citep{ghez_etal03,salim_gould99}.

Of course, the measurement of $\vhalo$ depends upon the assumption
that the halo rotation curve is flat; line-of-sight velocities would
be unaffected by adding a solid-body component to the angular velocity
of the halo, although proper motions would be (see below).

These are the main results of our study.  It is likely that
more science could be extracted from this sample, and completion of
the SDSS will allow it to be enlarged.  For the present, we simply
highlight certain interesting features of the data that have not yet
been thoroughly explored.

\begin{figure}
\includegraphics[width=\textwidth]{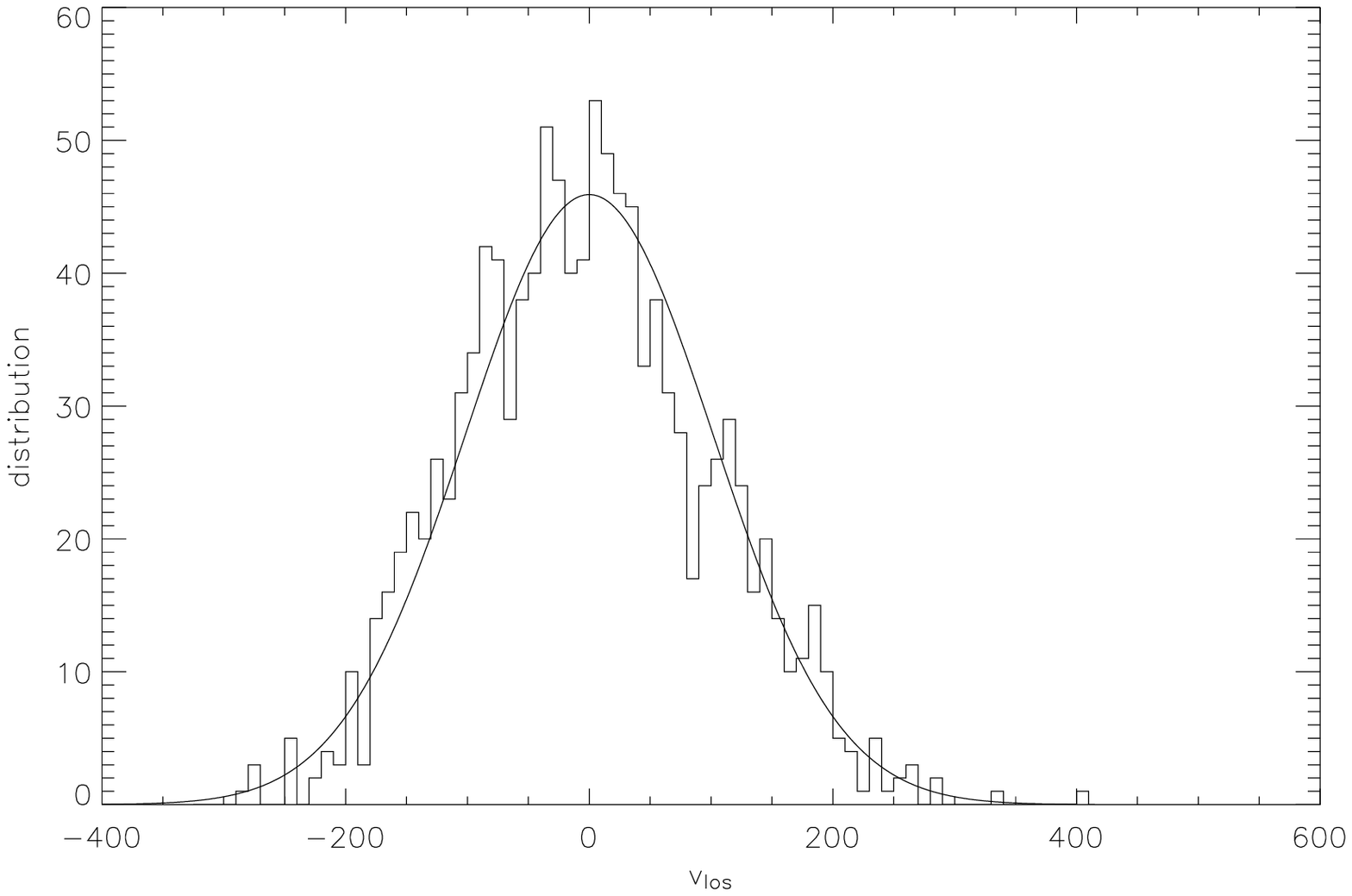}
\caption{Histogram of 
$\vlos = v_r + \mathbf{x} \cdot \vsun$ 
($\vsun = (10,5,7) \kms$ \citep{dehnen_binney_1998}),
with a gaussian of width $\sigmalos = \sigmalosval$.}
\label{f:vlos_dist}
\end{figure}

The distribution of $\vlos$ is plotted in
figure~\ref{f:vlos_dist} and beautifully follows a gaussian of width
$\sigmalos = \sigmalosval \pm \sigmaloserror \kms$.
The uncertainty quoted here is 
one-sigma statistical, to which should be added a similar
systematic uncertainty pending better quantification of our
measurement errors.  An estimate of this systematic error is easily
determined as $\sigmalos - \sqrt{\sigmalos^2 - \Delta v^2} \approx 3.4 \kms$
($\Delta v = 26 \kms$ from Paper~I).
The line-of-sight velocities (figure~\ref{f:vlos_dist}) are about as
gaussianly distributed as the number of stars would allow.  This is
consistent with (although not required by) the isotropy of the
velocity ellipsoid that we find, since a strongly anisotropic
ellipsoid, even if gaussian along each individual axis, would not
produce an exact gaussian in an average over many lines
of sight.  Gaussianity could result from large observational errors,
regardless of the intrinsic velocity distribution, if the errors were
gaussian, but our errors seem not to be large enough to dominate the
observed distribution (Paper~I).  By contrast, the
distribution of transverse velocities of halo stars in the solar
neighborhood is distinctly nongaussian, with a negative kurtosis
\citep{popowski_gould_1998}.  This is another indication that
the local halo differs from the more distant halo probed by our
BHB sample.
 
Another aspect of the data that could be thoroughly explored is the
smoothness of the BHB phase-space distribution.  The notion
of a stellar velocity ellipsoid implicitly presumes that phase
space is smoothly populated.  If instead the distribution is
very lumpy and the number of important lumps or streams is small,
then statistical characterizations of \emph{individual} halos, in
particular that of the Galaxy, may not be meaningful.  In cosmology,
however, statistical descriptions are unavoidable.
Current theory
holds that galactic halos have been built up by accretion of lesser
systems \citep{searle_zinn_1978,bullock_kravtsov_weinberg_2001}.
Accreted gas joins galactic disks, and the details of
its history are effaced by dissipation.  Stars that form in
these systems before they are accreted (and presumably also their
dark matter), however, behave collisionlessly and preserve a
partial record of their origins.  The structure of the stellar
and dark-matter halos should be related.
\citet{helmi_white_springel_03} conclude from a detailed analysis
and scaling of cosmological simulations that the number of
streams composing a halo similar to that of the Galaxy
should be large,
amounting to almost $10^4$ within $25\kpc$, and that most streams
joined the Galaxy early on, allowing more time for phase mixing.  The
number of our BHB stars is much smaller than this, so that one
might expect to find a smooth distribution.
And yet we see structure in our sample that is
consistent with the tidal stream of the Sagittarius Dwarf
(Paper~I), as previously delineated in SDSS data
\citep{ivezic_etal_2000,yanny_etal_2000,ibata_etal_2001a,
newberg_etal_2002} and in a survey of carbon stars
\citep{ibata_etal_2001b}.  Most of our results for the halo velocity
ellipsoid are scarcely affected by excluding that part of the sample
that lies close to the Sagittarius stream on the sky. 
In short, it is not yet clear whether
the kinematics of the Galactic halo admit a statistical decription.
Larger samples of stars will be necessary to clarify this.

In Paper~I, we used proper motions only as a check on the purity of our
sample.  Calibration against the QSOs shows that the proper motions of
individual BHB stars in our sample are smaller than measurement errors
by factors $\sim 3-5$, but statistical detection of the kinematics of
the population as a whole should certainly be possible.  
Among other
benefits, this would allow a direct constraint on the solid-body
component of halo rotation and a further test of the isotropy of the
velocity dispersions.  Systematic errors in the proper motion data
present many complexities, however, so we defer their analysis to a
later paper.  The quality of the USNO-B data should be appreciably
better than that of the USNO-A, and one should probably use the former
to determine the proper motions of SDSS objects.

Our BHB radial velocity data could be used to constrain the mass of the outer
Galaxy.  We have not yet attempted a serious dynamical analysis,
partly in order to do full justice to the kinematics, but also because
mass measurements rely more heavily on aspects of the data that are
less certain: namely, the photometric distances and the completeness
as a function of magnitude.  
However, since no kinematic information
was used in the selection of the sample, we may assume that the
velocities fairly represent the whole population of BHB stars, whose
spatial density is often described as a powerlaw in Galactocentric
radius, $n\propto \scriptr^{-\alpha}$, with exponent $\alpha\approx 3.5$
\citep{kinman_suntzeff_kraft_1994}.  
In that case, to the extent that the dispersions
are isotropic and constant with radius and rotation is negligible, 
and that the potential is spherical, Jeans' equations imply
\begin{eqnarray}\label{e:Jeans}
v_{\rm circ} &\approx&
187\left(\frac{\alpha}{3.5}\right)^{1/2}\left(\frac{\sigma}{100\kms}\right),\\
M(\scriptr) &\approx&
2.0\times 10^{11}\left(\frac{v_{\rm circ}}{187\kms}
\right)^{2}\left(\frac{\scriptr}{25\kpc}\right)\,M_\odot.
\end{eqnarray}
The value of $v_{\rm circ}$ so obtained, with $\sigma \approx 100 \kms$,
is substantially lower than the circular velocity in the vicinity
of the Sun.  Faced with similar evidence drawn from their own
independent BHB sample, \cite{Sommer-Larsen_etal97} concluded that the
tangential velocity dispersion of the halo probably rises beyond
$\scriptr\sim 20\kpc$ so as to exceed the radial dispersion and be consistent
with a constant circular velocity.  At such distances, however, the
tangential dispersions are very poorly constrained by line-of-sight
velocities without \emph{a priori} assumptions about the potential
(see Table~\ref{t:results}).  Rather than make such assumptions, one
may prefer to await further direct evidence.

\acknowledgements
We thank A.~Gould, B.~Paczy\'nski, R.~Lupton, and
J.~Hennawi for helpful discussions, and M.~Strauss for a close reading
of a draft of this paper.
GRK is grateful
for generous research support from Princeton University and
from NASA via
grants NAG-6734 and NAG5-8083.
Funding for the creation and distribution of the SDSS Archive has been
provided by the Alfred P.
Sloan Foundation, the Participating Institutions, the National Aeronautics
and Space Administration, the National Science Foundation, the U.S.
Department of Energy, the Japanese Monbukagakusho, and the Max Planck Society.  The SDSS Web
site is {\tt http://www.sdss.org/}.  The SDSS is managed by the
Astrophysical Research Consortium (ARC) for the Participating Institutions.
The Participating Institutions are The University of Chicago, Fermilab,
the Institute for Advanced Study, the Japan Participation Group, The Johns
Hopkins University, Los Alamos National Laboratory, the Max-Planck-Institute
for Astronomy (MPIA), the Max-Planck-Institute for Astrophysics (MPA),
New Mexico State University, the University of Pittsburgh, Princeton
University, the United States Naval Observatory, and the University of
Washington.

\appendix
\section{The maximum likelihood technique}\label{s:maxlike}
In this section we outline the maximum likelihood procedure used to
determine the velocity ellipsoid of the Galactic halo.

Halo stars are assumed to have a gaussian velocity distribution
whose principal axes align with spherical or cylindrical coordinates.
In this appendix we assume the velocity of the Sun relative to the 
galactic center is known and subtracted from stars' radial velocities, 
and that every star's position in space is known exactly; see
\S\ref{s:ve_overview} for a discussion of magnitude and velocity
errors.
We define $\mathbf{x}$ as the normalized heliocentric 
line-of-sight vector to the star, 
so that the line-of-sight velocity $\vlos = \mathbf{x} \cdot \mathbf{v}$ 
where $\mathbf{v}$ is the true space velocity of the star.
The three components of each of the vectors 
$\mathbf{x}$, $\mathbf{v}$, and $\boldsymbol{\sigma}$
should be thought of as projections along the principal axes of the
velocity ellipsoid at the specific point in space.
The probability $\phi\,d\vlos$ of observing a given star with a radial
velocity between $\vlos$ and $d\vlos$ is expressed as an integral
over the possible combination of true space velocities that
would result in an observation between $\vlos$ and $d\vlos$:
\begin{equation}
\phi(\vlos)\,d\vlos = \int_{-\infty}^{\infty}\int_{-\infty}^{\infty}
	 \phi_1(v_1) \phi_2(v_2) \phi_3(v_3) \, dv_1 dv_2 dv_3
\end{equation}
where $v_1$, $v_2$, and $v_3$ are the components of the star's velocity
in the directions of the principal axes.  Here
$\phi_i(v_i)\,dv = (2\pi\sigma_i^2)^{-1/2} \exp{(-v_i^2 / 2\sigma_i^2)}\,dv$ 
where $\sigma_i$ is the
component of the velocity ellipsoid in the $i$ direction.
Geometry dictates a relation between $d\vlos$ and the other infinitesimal
quanitites: $d\vlos = |x_i| dv_i$.  Also, the vector $\mathbf{v}$
is constrained by $\vlos$ by the relation 
$\vlos = \mathbf{x} \cdot \mathbf{v}$ so finally our integral becomes
\begin{equation}
\phi(\vlos)\,d\vlos = \frac{d\vlos}{|x_1|} 
	\int_{-\infty}^{\infty}\int_{-\infty}^{\infty} dv_2 dv_3 \, 
	\phi_1\left(\frac{1}{x_1} (\vlos - x_2 v_2 - x_3 v_3) \right)
	\phi_2(v_2) \phi_3(v_3)
\end{equation}
Here the integral integrates over the $2$ and $3$ directions, which
is mathematically correct if $x_1 \ne 0$, but to avoid small-number
numerical problems, in practice the
code shuffles the indices as appropriate to ensure that the largest
component of $\mathbf{x}$ is placed in the denominator.
The value of the integral is
\begin{eqnarray}\label{e:probability}
\phi(\vlos) &=& \frac{1}{\mathit{\Gamma}} 
	\exp{\left(\frac{-\vlos^2}{2 \sigmaloshat^2}\right)} \\
\frac{1}{2 \sigmaloshat^2} = C' - B'^2/4A' \qquad 
\mathit{\Gamma} &=& |x_1| \sigma_1 \sigma_2 \sigma_3 \sqrt{8\pi A A'} \nonumber\\
A' = -\frac{x_3^2}{2 \sigma_1^2 x_1^2} (D - 1) +
	\frac{1}{2 \sigma_3^2} \qquad
B' &=& \frac{x_3}{\sigma_1^2 x_1^2} (D - 1) \qquad
C' = -\frac{1}{2 \sigma_1^2 x_1^2} (D - 1) \nonumber \\
A = \frac{1}{2 \sigma_2^2} + 
	\frac{x_2^2}{2 \sigma_1^2 x_1^2} \qquad 
D &=& \frac{\sigma_2^2 x_2^2}{\sigma_1^2 x_1^2 + \sigma_2^2 x_2^2}. \nonumber
\end{eqnarray}
Note that at a given position, stars are most likely to be found 
at zero velocity 
(with respect to the galactic center) but have a characteristic
velocity dispersion $\sigmaloshat$, which is a function of the geometry of
the stars with respect to the Sun and of $\boldsymbol{\sigma}$.
Since the probability of observing a set of stars with known space 
locations and observed line-of-sight velocities is the product
of the individual probabilities, the log-likelihood is
\begin{equation}\label{e:loglike}
\loglike(\boldsymbol{\sigma}) = 
	-\frac{N}{2} \ln{\left(8\pi \sigma_1^2 \sigma_2^2 \sigma_3^2 \right)} +
	\sum{\left[\left(\frac{B'^2}{4A'} - C'\right)\vlos^2 - 
		\frac{1}{2} \ln{(x_1^2 A A')}\right]}
\end{equation}
where the summation is over all observed stars.  Assuming a uniform prior
\citep{lupton_1993}, we find the maximum of 
$\loglike(\sigma_1, \sigma_2, \sigma_3)$ to determine the most likely values
of the velocity ellipsoid $\boldsymbol{\sigma}$.

If quantities such as the velocity of the Sun with respect to the
Galactic center or the bulk rotation velocity of the halo are 
free parameters, they can be determined from this maximum-likelihood
estimate.  These quantities affect $\vlos$ for every star, so it is
straightforward to extrapolate equation~\ref{e:loglike} to 
$\loglike = \loglike(\boldsymbol{\sigma}, \vsun, v_{\rm{halo}})$.

\clearpage


\clearpage


\clearpage

\begin{deluxetable}{ccccll}
\tablecaption{Recent results for the local halo velocity ellipsoid.
The velocity ellipsoid components $(R,\phi,z)$ correspond to galactocentric
cylindrical coordinates and point radially (outwards), retrograde to galactic
rotation, and towards the north galactic pole, respectively.  Entries
in the fourth column assume $\vlsr = (0,220,0) \kms$ and 
$\vsun - \vlsr = (10,5,7) \kms$.
	\label{t:localhalo}}
\rotate
\tablewidth{0pt}
\tablehead{\colhead{$\sigma_R$}&\colhead{$\sigma_\phi$}&\colhead{$\sigma_z$}&
	\colhead{$\vhalo$}&\colhead{comments}&\colhead{source}
}
\startdata
   $125\pm 11$ &  $96\pm 9$ & $88\pm 7$ & --- & $\metal<-1.2$; $V\le 11$~mag &
	\citet{Norris_etal85}\\
$131\pm6$ & $106\pm6$ & $85\pm4$ & $32\pm10$ & $\metal<-1.2$ & \citet{norris_1986}\\
$154\pm18$ & $102\pm27$ & $107\pm15$ & $7\pm23$ & $\metal<-1.5$; red giants &
     \citet{carney_latham_1986}\\
$133\pm8$ & $98\pm13$ & $94\pm6$ & $21\pm15$ & $D\sim4\kpc$, $\metal<-1.6$ & 
	\citet{Morrison_etal90}\\
$145\pm10$ & $100\pm10$ & --- & $30\pm10$ & $\metal<-1.5$; kin. sel. subdwarfs &
        \citet{Ryan_Norris91}\\
 --- & --- & $54\pm4$ & --- & metal-poor RGs, $D<4\kpc$ &
        \citet{Ratnatunga_Yoss91}\\
$168\pm13$ & $102\pm8$ & $97\pm7$ & $15\pm12$ & RR Lyr: stat-$\pi$ \&
        $\vlos$ & \citet{layden_etal_1996}\\
$151\pm9$ & $121\pm7$ & $103\pm6$ & $13\pm10$ & $|Z|<4\kpc$, $\metal<-2.2$ & 
	\citet{Chiba_Beers00}\\
$162\pm1$ & $106\pm2$ & $89\pm2$ & --- & stat $\pi$,
                  $\langle V\rangle-V_\odot\equiv-216.6\kms$ & \citet{Gould03}\\
\enddata
\end{deluxetable}

\begin{deluxetable}{ccccccccccc}
\tablecaption{Results \label{t:results}}
\rotate
\tabletypesize{\scriptsize}
\setlength{\tabcolsep}{0.01in}
\tablewidth{0pt}
\tablehead{
\colhead{criteria} & \colhead{$N_{\rm{stars}}$} & \colhead{c.s.} &      
\colhead{errors} &
\colhead{$v_{\odot,x}$} & \colhead{$v_{\odot,y}$} & \colhead{$v_{\odot,z}$} &
\colhead{$v_{\rm{halo}}$} &
\colhead{$\sigma_r$ (sph), $\sigma_R$ (cyl)} & 
\colhead{$\sigma_\theta$ (sph), $\sigma_\phi$ (cyl)} & 
\colhead{$\sigma_\phi$ (sph), $\sigma_z$ (cyl)} }
\startdata
 all & 1170 & sph & 0 & 10.0 & 225.0 & 7.0 & 0.0 & 101.4 $\pm$ 2.8(2.5) & 97.7 $\pm$ 16.4(14.5) & 107.4 $\pm$ 16.5(13.9) \\
 all & 1170 & sph & 0 & 10.0 & 225.0 & 7.0 & -6.0 $\pm$ 10.2(10.2) & 101.4 $\pm$ 2.8(2.5) & 97.9 $\pm$ 16.6(14.5) & 107.9 $\pm$ 16.6(13.9) \\
 all & 1170 & sph & 0 & 4.8 $\pm$ 6.8(6.5) & 245.9 $\pm$ 13.6(6.7) & -7.2 $\pm$ 4.0(3.8) & 23.8 $\pm$ 20.2(10.1) & 100.6 $\pm$ 2.8(2.4) & 94.8 $\pm$ 16.6(14.6) & 108.2 $\pm$ 16.5(13.8) \\
 all & 1170 & sph & m & 10.0 & 225.0 & 7.0 & 0.0 & 101.5 $\pm$ 2.8(2.5) & 102.1 $\pm$ 18.0(15.1) & 103.5 $\pm$ 18.2(14.7) \\
 all & 1170 & sph & m/v & 10.0 & 225.0 & 7.0 & 0.0 & 96.9 $\pm$ 3.0(2.6) & 97.5 $\pm$ 18.8(15.8) & 99.1 $\pm$ 19.0(15.3) \\
 all & 1170 & cyl & 0 & 10.0 & 225.0 & 7.0 & 0.0 & 98.7 $\pm$ 7.7(5.7) & 104.7 $\pm$ 16.8(14.1) & 102.8 $\pm$ 5.2(3.3) \\
 all & 1170 & cyl & m & 10.0 & 225.0 & 7.0 & 0.0 & 99.2 $\pm$ 7.8(5.7) & 102.1 $\pm$ 17.9(14.8) & 102.9 $\pm$ 5.2(3.3) \\
 all & 1170 & cyl & m/v & 10.0 & 225.0 & 7.0 & 0.0 & 94.5 $\pm$ 8.1(6.0) & 97.6 $\pm$ 18.7(15.5) & 98.5 $\pm$ 5.4(3.5) \\
\cline{1-11} \\
 $g<18$ & 733 & sph & 0 & 10.0 & 225.0 & 7.0 & 0.0 & 99.4 $\pm$ 4.3(3.4) & 100.0 $\pm$ 17.2(14.7) & 110.5 $\pm$ 17.1(13.7) \\
 $g<18$ & 733 & cyl & 0 & 10.0 & 225.0 & 7.0 & 0.0 & 97.2 $\pm$ 9.6(7.6) & 109.1 $\pm$ 18.0(13.8) & 100.9 $\pm$ 6.5(4.3) \\
\cline{1-11} \\
 $g<16$ & 227 & sph & 0 & 10.0 & 225.0 & 7.0 & 0.0 & 105.4 $\pm$ 10.9(8.2) & 85.7 $\pm$ 16.9(14.6) & 117.6 $\pm$ 24.5(18.2) \\
 $g<16$ & 227 & cyl & 0 & 10.0 & 225.0 & 7.0 & 0.0 & 108.8 $\pm$ 19.7(16.2) & 122.0 $\pm$ 23.8(17.5) & 94.1 $\pm$ 12.5(8.3) \\
\cline{1-11} \\
 $g>18$ & 437 & sph & 0 & 10.0 & 225.0 & 7.0 & 0.0 & 103.2 $\pm$ 6.6(3.6) & 28.6 $\pm$ 1013.4(741.5) & 113.8 $\pm$ 211.2(140.1) \\
 $g>18$ & 437 & cyl & 0 & 10.0 & 225.0 & 7.0 & 0.0 & 99.7 $\pm$ 13.4(8.6) & 113.3 $\pm$ 207.5(142.0) & 104.5 $\pm$ 10.4(5.5) \\
\cline{1-11} \\
 excl. Sgr & 775 & sph & 0 & 10.0 & 225.0 & 7.0 & 0.0 & 100.6 $\pm$ 3.6(3.0) & 89.9 $\pm$ 34.2(26.9) & 106.8 $\pm$ 19.5(14.7) \\
 excl. Sgr & 775 & cyl & 0 & 10.0 & 225.0 & 7.0 & 0.0 & 96.3 $\pm$ 9.5(7.1) & 102.1 $\pm$ 18.8(15.2) & 102.6 $\pm$ 6.3(4.1) \\
\enddata
\tablecomments{
The criteria used to select subsamples from the entire
sample of $\nbhb$ BHB stars are listed in the first column.  The next
three columns indicate the number of stars in the subsample, the
coordinate system used, and whether magnitude or velocity errors were
considered in the maximum likelihood analysis.
The quoted errors not in parentheses are
considering correlations among the parameters; the errors in parentheses
do not consider correlations.}
\end{deluxetable}

\begin{deluxetable}{ccrrrcrrr}
\tablecaption{Mean and rms of determinations for
$\boldsymbol{\sigma}$ over 100 Monte Carlo simulations for four cases
\label{t:mc_meanrms}}
\tablewidth{0pt}
\tablehead{
\colhead{} & \colhead{} & 
\multicolumn{3}{c}{mean} & \colhead{} & \multicolumn{3}{c}{rms} \\
\cline{3-5} \cline{7-9} \\
\colhead{error} & \colhead{color in figure~\protect{\ref{f:ve_mc}}} 
	& \colhead{$\sigma_r$} & 
	\colhead{$\sigma_\theta$} & \colhead{$\sigma_\phi$} & 
	\colhead{} & \colhead{$\sigma_r$} & 
	\colhead{$\sigma_\theta$} & \colhead{$\sigma_\phi$} }
\startdata
0 & black     & 150.2 &  98.0 &  73.7 & &   3.4 &  19.7 &  35.6 \\
m & blue      & 149.5 &  98.7 &  73.7 & &   3.9 &  18.3 &  35.4 \\
v & red       & 152.3 & 101.1 &  79.3 & &   3.8 &  19.5 &  35.9 \\
m/v/c & green & 146.6 &  96.8 & 104.7 & &   4.0 &  20.5 &  30.7 \\
\enddata
\tablecomments{The type of error(s) considered is indicated in the
leftmost column: ``0,'' ``m,'' ``v,'' and ``c'' refer to no error,
magnitude error, velocity error, and contamination, respectively.
Compare the mean values of $\boldsymbol{\sigma}$ to the input
to the simulation of $(150,100,80) \kms$.}
\end{deluxetable}

\end{document}